\documentclass[twocolumn,showpacs,prb]{revtex4}
\usepackage{amssymb}

\usepackage{graphics}
\usepackage{epsfig,epsf}
\usepackage{dcolumn}
\usepackage{amsmath}

\begin{document}
\title{Boson-assisted tunneling in layered metals }
\author{D. B. Gutman and D. L. Maslov}
\affiliation{Department of Physics, University of Florida, Gainesville,FL 32611, USA}

\date{\today}

\begin{abstract}
A theory for boson-assisted tunneling via randomly distributed
resonant states in a layered metals is developed. As particular
examples, we consider the electron-phonon interaction and the
interaction between localized and conduction electrons. The theory
is applied to explain a non-monotonic variation of the out-plane
resistivity with temperature observed in quasi-two-dimensional
metals.
\end{abstract}
\maketitle

\section{Introduction}

\label{sec:intro}

Electron transport in layered metals exhibits a qualitatively different
behavior of the in-plane $\left( \rho _{ab}\right) $ and out-of-plane ($\rho
_{c})$ resistivities: whereas the temperature dependence of $\rho _{ab}$ is
metallic-like, that of $\rho _{c}$ is either insulating-like or even
non-monotonic. This behavior is observed in various materials, such as
high-temperature superconductors \cite{ginsberg}, sodium and bismuth
cobaltate oxides \cite{Terasaki,Loureiro,Tsukada,Valla}, the layered perovskite
Sr$_{2}$RuO$_{4}$\cite{maeno94,hussey,srruo}, dichalogenides \cite{frindt},
graphite \cite{graphite}, organic metals \cite{organics}, and other systems.
It is quite remarkable that the $c$-axis resistivity behaves similarly in
materials with otherwise drastically different properties, ranging from
weakly (graphite) to strongly (Sr$_{2}$RuO$_{4}$) correlated Fermi liquids
and then to non-Fermi liquids (HTC, cobaltate oxides), and with Fermi
energies spanning the interval from a few eV (in most layered metals) to $25$
meV (in graphite). Also, the magnitude of $\rho _{c}$ varies from a few m$%
\Omega \cdot $cm (e.g., in Sr$_{2}$RuO$_{4})$ to a few $\Omega \cdot $cm
(e.g., in organics and (Bi$_{1-x}$Pb$_{x}$)$_{2}$Sr$_{3}$Co$_{2}$O$_{3})$.
The great variety of systems and associated energy scales suggests
that the origin of this effect is not related to the specific properties of
any of the compounds but rather to what they have in common, i.e., strong
anisotropy.

At the level of non-interacting electrons, layered systems are metals with
strongly anisotropic Fermi surfaces. A simple but instructive model is that
of free motion along the layers and nearest-neighbor hopping between the
layers. In this model, the single-particle spectrum is
\begin{equation}
\epsilon _{\mathbf{k}}=\mathbf{k}_{||}^{2}/2m_{ab}+2J_{c}\left( 1-\cos
k_{z}d\right) ,  \label{spectrum}
\end{equation}
where $\mathbf{k}_{||}$ and $k_{z}$ are in the in-plane and $c$-axis
components of the momentum, respectively, $m_{ab}$ is the in-plane mass, $J_{c}$
is the hopping matrix element in the $c$-axis direction, and $d$ is the
lattice constant in the same direction. Strong mass anisotropy is guaranteed
by the condition $m_{ab}\ll m_{c}=1/2J_{c}d^{2}$.(We set $\hbar=k_B=1$ through the
rest of the paper.) For $E_{F}<2J_{c}$ the
Fermi surface is closed (as it is in graphite); for $E_{F}>2J_{c}$ the Fermi
surface is open (as it is in the majority of layered materials). For weakly
coupled layers ($E_{F}\gg J_{c}),$ the equipotential surfaces are
``corrugated cylinders'' with slight modulation along the $c$-axis (see
Fig.1).

Transport in metals is commonly described via the Boltzmann equation for the
distribution function of electrons ($f)$ in the phase space. To study the
linear response conductivity, it is sufficient to consider a weak electric
field. In this case, $f=f_{0}+f_{1}$, where the non-equilibrium correction $%
f_{1}$ to the Fermi function $f_{0}$ satisfies
\begin{equation}
\frac{\partial f_{1}}{\partial t}+\mathbf{v\cdot \nabla }_{\mathbf{r}%
}f_{1}+I[f_{1}]=-\frac{\partial f_{0}}{\partial \epsilon }e\mathbf{v\cdot }%
\mathbf{E}\,.
\end{equation}
Here $I$ is a linearized collision integral and $\nolinebreak {\mathbf{v}%
=\nabla _{\mathbf{k}}\epsilon _{\mathbf{k}}}$. The tensor of the $dc$ electrical
conductivity is given by
\begin{equation}
\sigma _{\alpha \beta }=e^{2}\int \frac{d^{3}p}{\left( 2\pi \right) ^{3}}%
\left( -\frac{\partial f_{0}}{\partial \epsilon }\right) \mathbf{v}_{\alpha
}(\mathbf{p})\hat{I}^{-1}\mathbf{v}_{\beta }(\mathbf{p})\,,
\end{equation}
where the integration goes over the Brillouin zone. 
For all known types of the inelastic interaction (electron-phonon,
electron-electron, electron-magnon, etc.), the operator $I^{-1}$ decreases
with the temperature; thus all components of the conductivity tensor must
decrease with $T$ as well (although the particular forms of the $T$
dependence may be different for different components). This is not what the
experiment shows.
\begin{figure}[tbp]
\begin{center}
\epsfxsize=0.6\columnwidth
\epsfysize=0.7\columnwidth
 \epsffile{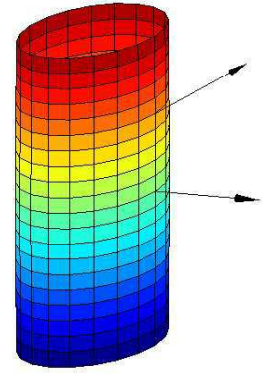}
\end{center}
\caption{Fermi surface corresponding to Eq.(\ref{spectrum}) with Fermi
velocity vectors at two different points.}
\label{fig:FS}
\end{figure}



The situation when Bolztmann equation fails so dramatically is rather
unusual, given that other examples of its breakdown, e.g., weak localization
and Altshuler-Aronov effects in the conductivity, indicate some non-trivial
quantum interference effects. The fact that this breakdown occurs only in the
$c$-direction, makes the situation even more puzzling. A vast amount of
literature, addressing various aspects of this problem, is accumulated at
this point. The proposed models can be roughly divided in two groups. The
first group is trying to find the explanation of the anomalous $c$-axis
transport within a model of an anisotropic metal in which an electron
interacts with potential impurities, phonons, and/or other electrons. The
breakdown of the Boltzmann equation is associated either with Anderson
localization in the $c$ (but not in the in-plane) direction or from the
``coherent-incoherent crossover'', which is believed to occur when the
tunneling time between the layers ($J_{c}^{-1}$) becomes longer than the
inelastic time ($\tau _{\text{in}}$) \cite{kumar} or the thermal time ($%
T^{-1}$) \cite{anderson_criterion}, i.e., when $J_{c}\min \{\tau _{\text{in}%
},T^{-1}\}\lesssim 1.$ However, as far as Anderson localization is
concerned, it was shown that it occurs only for exponentially small $J_{c}$ [%
$J_{c}\lesssim \tau _{\text{el}}^{-1}\exp \left( -E_{F}\tau _{\text{el}%
}\right) ,$ where $\tau _{\text{el }}$ the elastic scattering time] and then
only simultaneously in all directions \cite{woelfle,abrikosov_loc,dupuis}.
The breakdown of the Boltzmann equation due to the coherent-incoherent
crossover for the electron-phonon interaction has been reconsidered in our
recent paper \cite{GM}, where we have shown that the Prange-Kadanoff
derivation of the Boltzmann equation \cite{prange} applies equally well to
the anisotropic case. The only condition for the validity of the Boltzmann
equation for both isotropic and anisotropic cases is the large value of the
``dimensionless conductance'' $E_{F}\tau _{\text{in}}$, where
$\tau _{\text{in}}$ is the inelastic scattering time, regardless of whether
$J_{c}\min \{\tau _{\text{in}},T^{-1}\}$ is large or small, provided that
the Migdal parameter $s/v_{F}$ ($s$ is the sound velocity) is small over the
entire Fermi surface, which is the case for Fermi surfaces of the type shown
in Fig. \ref{fig:FS}. This argument can be readily extended to any situation
when the self-energy is local, i.e., independent of the electron's momentum,
and applies not only to Fermi liquids but also to non-Fermi liquids.
Therefore, for a large class of non-perturbative interactions the
coherent-incoherent crossover does not occur, and the $c$-axis resistivity
is supposed to maintain its metallic character. On the experimental side,
the existence of the coherent-incoherent crossover has been questioned by a
recent observation of angular magneto resistance oscillations in a layered
organic metal \cite{singleton} well above the temperature of the expected
crossover. Moreover, the coherent-incoherent crossover on its own does not
account for a non-metallic conductivity, as in this scenario the
conductivities in both the coherent and incoherent regimes are proportional
to the same scattering time \cite{ioffe},\cite{mckenzie} and thus exhibit
a metallic $T$ dependence.

Along a similar line of reasoning, it was suggested that the non-metallic $c$%
-axis transport is related to the Fermi-to-non-Fermi-liquid crossover
(understood as a smearing of the quasiparticle peak in the spectral
function), which occurs at high enough temperatures in some materials \cite
{Valla,Millis}. Indeed, the angle-resolved photoemission experiments on
sodium and bismuth cobaltate oxides show that the temperature at which the
quasiparticle peak is smeared is of the same order as temperature $T_{M}$
at which $\rho _{c}$ exhibits a maximum \cite{Valla,Millis}. Although this
coincidence is suggestive, it runs against the Prange-Kadanoff argument \cite
{prange} which shows that the existence of quasiparticles is \emph{not }a
pre-requisite for Boltzmann-like transport. Also, it is not clear at the
moment whether the relation between smearing of the quasi-particle peak and
the maximum in $\rho _{c}$ is common for all materials. For example, $T_{M}$
in the HOPG graphite is low enough ($\sim 40$ K) \cite{graphite}, so that
quasiparticles are still
well-defined at $T\sim T_M$. Finally, the model of Fermi-to-non-Fermi-liquid crossover on its
own does not account for the anomalous transport, especially given that the
in-plane resistivity shows no dramatic signs of this crossover.

Within the first group is also the polaron model \cite{polaron_mckenzie}
\cite{polaron_schofield}, which is capable of a quantitative description of
the experiment, given that polarons are stable. The latter assumption,
however, requires a very strong anisotropy of the phonon spectrum, which
needs to be more anisotropic than the electron one, which is not the case at
least in some representative materials, e.g., in Sr$_{2}$RuO$_{4}$ \cite
{elastic}. Finally, there is an explanation based on the zero-bias anomaly
in tunneling between the layers, resulting from a suppression of the
single-particle density of states of in-plane electrons, e.g., via a
pseudogap mechanism in high $T_{c}$ cuprates \cite{ioffe},\cite{Turlakov}.
However, the zero-bias mechanism is only valid on the incoherent side of the
coherent-incoherent crossover which, as we argued earlier, does not occur
for a large class of interaction.

The second group assumes that the primary reason of the anomalous behavior
of the $c$-axis resistivity is transport through inter-plane defects. The
temperature dependence is introduced either phenomenologically \cite{sauls},
\cite{levin} or through thermal occupation of electron states in the
conducting layers \cite{frindt,uher,abrikosov_res}, or else, as it was done in the short version of this paper \cite
{GM}, through a phonon-assisted tunneling. Whether inter-plane disorder is
the reason for the anomalous $c$-axis transport in all layered materials is
not clear at the moment. Currently, there is a number of observations
pointing at the role of inter-plane disorder as a mediator of $c$-axis
transport. For example, recent experiment \cite{Analytis} has shown that the
radiation damage of an organic metal \emph{reduces} rather then increases $%
\rho _{c}$ at temperatures around $T_{M}.$ Also, variability of $\rho _{c}$
in graphite samples prepared in different ways (including the natural ones)
provides an indirect argument for the role of inter-plane disorder. It seems
worthwhile to explore the consequences of a model involving inter-plane
disorder and this what we will do in this paper.

We assume that, in addition to normal impurities, a layered crystal also
contains a number of resonant impurities located in the inter-planar space.
An electron moves between the layers via two mechanisms: the first one is
direct tunneling, augmented by scattering at normal (non-resonant)
impurities, phonons, etc., and the second one is resonant tunneling through
defect sites. Direct tunneling between the layers forms a band state smeared
by various scattering processes, which include the non-resonant part of
scattering by inter-plane defects. Transport of these states is described
by the Boltzmann conductivity $\sigma _{c}^{B}$ which has a metallic
temperature dependence. As resonant tunneling opens a new channel of
conduction, the total conductivity can be described by a phenomenological
formula \cite{levin,hussey,Analytis}
\begin{equation}
\sigma _{c}=\sigma _{c}^{B}+\sigma _{c}^{\text{res}},  \label{cond}
\end{equation}
where $\sigma _{c}^{\text{res}}$ is the resonant-impurity contribution.On the other hand, the
in-plane conductivity remains largely unaffected by inter-plane disorder, so
that $\sigma _{ab}=\sigma _{ab}^{B}.$ Regardless of a particular tunneling
mechanism, the resonant part $\sigma _{c}^{\text{res}}$ increases with $T$.
Consequently, the band channel, which is weak to begin with due to a small
value of the inter-plane transfer $J_{c},$ is short-circuited by the
resonant one at high enough temperatures. Accordingly, $\sigma _{c}$ goes
through a minimum at a certain temperature (and $\rho _{c}=\sigma _{c}^{-1}$
goes through a maximum). More generally, $\rho _{c}$ may exhibit a variety
of behaviors , discussed in Sec.\ref{comparison} 

In this paper, we develop a microscopic theory of resonant tunneling through
a wide band of energy levels positioned randomly in between two conducting
layers and coupled to a fluctuating field of bosonic excitations (Sec.\ref{sec:gen}). Two
particular examples of such a field, considered here, are phonons and the
dynamic Coulomb field of all conduction electrons in a crystal. The case of
phonon-assisted tunneling through a single junction was considered before in
Refs.~[\onlinecite{Glazman_1988,wingreen}]. Our formalism reproduces the
general results of Refs.~[\onlinecite{Glazman_1988,wingreen}], as well as the
low$-T$ behavior of the conductivity, found in Ref.~[\onlinecite{Glazman_1988}]. In Sec.~\ref{sec:eph}, we
obtain a detailed, non-perturbative expression for $\sigma _{c}^{\text{res}%
}$ for the electron-phonon interaction and showed that $\sigma _{c}^{\text{%
res}}$ saturates at temperatures higher than

\begin{equation}
T_{s}=\lambda \omega _{D}  \label{ts}\,,
\end{equation}
where $\omega _{D}$ is the Debye frequency and $\lambda $ is the
dimensionless coupling constant for on-site electrons. This prediction is
important for discriminating the phonon-assisted mechanism against other
effects. A strong on-site Coulomb interaction was shown to result in a
Kondo anomaly in the tunneling conductance \cite{Glazman_Raikh,Ng}.
The effect of the Coulomb interaction between on-site and conduction
electrons was considered by Matveev and Larkin \cite{Larkin_Matveev} in the
context of a threshold singularity in the non-linear current-voltage
characteristic, observed in Ref.~[\onlinecite{Geim}]. The new element of this work is
that we consider the effect of the Coulomb interaction between on-site and
conduction electrons on the linear tunneling conductance both in the
ballistic and diffusive regimes of conduction electrons' motion (sec.~\ref{sec:ee}). The most
interesting result of this analysis is the scaling behavior $\sigma _{c}^{%
\text{res}}\propto T^{\eta_{B} }$ in the ballistic regime, where $\eta_{B} $ is the
dimensionless coupling constant for the Coulomb interaction. In Sec.\ref{sec:discussion}, we analyze the
results. In Section \ref
{comparison}, we compare our theory to the experiment and show that equation
(\ref{cond}) describes well the non-monotonic $T$-dependence of $\rho _{c}$
in Sr$_{2}$RuO$_{4}$ \cite{srruo}and $\kappa $-(BEDT-TTF)$_{2}$Cu(SCN)$_{2}$
\cite{Analytis}. Our conclusions are given in Sec.\ref{conclusions}.

\section{Boson-Assisted tunneling}

\subsection{General formalism}
\label{sec:gen}
We consider tunneling through a single resonant impurity located in between
two metallic layers. To account for a finite concentration of such
impurities, we will average the result with respect to the positions of the
resonant center in real space and also within the energy band. In doing so,
we neglect the effects of interference between different resonant centers,
as well as between resonant and non-resonant scattering. Tunneling through
more than one impurity was considered theoretically in %
\onlinecite{Glazman_Matveev} and observed experimentally in thick ($>1$ nm)
tunneling junctions \cite{Beasley}, but it is less likely to occur in
tunneling through thin ( $\lesssim 1$ nm) inter-plane spacings, so we will
disregard such a possibility here.

We use the tunneling Hamiltonian description

\begin{equation}
H=H_{\mathrm{a}}+H_{\mathrm{d}}+H_{\mathrm{c}}+H_{\mathrm{ac}}+H_{\mathrm{cd}%
}\,+H_{\mathrm{inel}}\,.
\end{equation}
The free part of the Hamiltonian
\begin{equation}
H_{a}=\sum_{\mathbf{k}_{||}}\epsilon _{\mathbf{k}_{||}}\hat{a}_{\mathbf{k}%
_{||}}^{\dagger }\hat{a}_{\mathbf{k}_{||}}\,\,,\,\,H_{d}=\sum_{\mathbf{k}%
_{||}}\epsilon _{\mathbf{k}_{||}}\hat{d}_{\mathbf{k}_{||}}^{\dagger }\hat{d}%
_{\mathbf{k}_{||}}\,.
\end{equation}
represents metallic layers. The tunneling part of the Hamiltonian
\begin{eqnarray}
&&H_{ac}=\sum_{k}g_{\mathrm{ca}}(\mathbf{k}_{||})(\hat{a}_{\mathbf{k}%
_{||}}^{\dagger }\hat{c}+\hat{c}^{\dagger }\hat{a}_{\mathbf{k}_{||}})\,, \\
&&H_{cd}=\sum_{\mathbf{k}_{||}}g_{\mathrm{cd}}(\mathbf{k}_{||})(\hat{d}_{%
\mathbf{k}_{||}}^{\dagger }\hat{c}+\hat{c}^{\dagger }\hat{d}_{\mathbf{k}%
_{||}})\,
\end{eqnarray}
describes hopping on and off the resonant site, located at point $\mathbf{r}%
_{i}=(\mathbf{r}_{||}=\mathbf{0},z_{i})$ in between the layers. Dynamics of
the resonant level is accounted by
\begin{equation*}
H_{\mathrm{c}}=(\epsilon _{0}+\hat{\phi}(\mathbf{r}_{i},t))\hat{c}^{\dagger }%
\hat{c}\,,
\end{equation*}
where the time-dependent operator $\hat{\phi}(\mathbf{r},t)$ describes the
fluctuations of the electrostatic potential. Although a general result can
be obtained for an arbitrary (but Gaussian) field $\hat{\phi},$ we will be
mostly interested in the case when this field arises due to fluctuations in
the positions of ions and electrons
\begin{equation}
\hat{\phi}(\mathbf{r},t)=\hat{\phi}^{\mathrm{ph}}(\mathbf{r},t)+\hat{\phi}^{%
\mathrm{e}}(\mathbf{r},t).
\end{equation}
The contribution arising from the crystal degrees of freedom is the
potential produced by the displacement wave
\begin{equation}
\hat{\phi}^{\mathrm{ph}}(\mathbf{r},t)=\sum_{\mathbf{q}}\alpha _{\mathbf{q}%
}(b_{\mathbf{q}}^{\dagger }e^{i\omega _{\mathbf{q}}t-i\mathbf{q\cdot r}}-b_{%
\mathbf{q}}^{-i\omega _{q}t+i\mathbf{q\cdot r}})\,,
\end{equation}
where $b_{\mathbf{q}}^{\dagger }$ is the phonon creation operator, $%
\nolinebreak {\alpha }_{\mathbf{q}}{\ }$is the vertex of the electron-phonon
interaction, and $\omega _{\mathbf{q}}$ is the phonon dispersion. The
electronic part of the potential is expressed through fluctuation of the
electron density $\hat{\rho}$ in the \emph{entire} crystal, including the
two layers accounted for in $H_{\mathrm{a}}$ and $H_{\mathrm{d}}$%
\begin{equation}
\hat{\phi}^{\mathrm{e}}(\mathbf{r},t)=\int d\mathbf{r}^{\prime }V_{0}(|%
\mathbf{r}-\mathbf{r}^{\prime }|)\hat{\rho}(\mathbf{r}^{\prime },t)\,,
\end{equation}
where $V_{0}(\left| \mathbf{r}\right| )=4\pi e^{2}/|\mathbf{r}|$.

We assume that the potential localizing the electron is of a short range and
therefore, when the site is empty, it does not affect the motion of
conduction electrons. This situation is different from that considered in
the context of the Fermi edge singularity in resonant tunneling \cite
{Mahan,Larkin_Matveev}, where the Coulomb field of the empty, charged site
scatters conduction electrons.
We also assume that the on-site Coulomb repulsion is so large that the
double occupancy of the resonant center is forbidden, and the spin degree of
freedom does not play any role in tunneling. The Kondo effect in tunneling
through a resonant impurity was considered in Refs.~[\onlinecite{Glazman_Raikh}].

The last term in the Hamiltonian, $H_{\mathrm{inel}},$ consists of the
free-phonon part
\begin{equation}
H_{\mathrm{ph}}=\sum_{\mathbf{q}}\omega _{\mathbf{q}}b_{\mathbf{q}}^{\dagger
}b_{\mathbf{q}}\,.
\end{equation}
and of the part describing the dynamics of \emph{all} mobile electrons in
the crystal. This dynamics controls the behavior of the density
fluctuations, $\hat{\rho}(\mathbf{r},t)\,.$

The effect of the electron-phonon interaction on tunneling is traditionally
referred to as phonon-assisted tunneling. By analogy, the effect of the
electron-electron interaction can be called electron-assisted tunneling. We
show that the problem of assisted tunneling allows for an exact solution for
any bosonic field $\hat{\phi}$, given its fluctuations in time are Gaussian.
Remarkably, the effect of different bosonic degrees of freedom (such as
phonons, plasmons, or diffusive density fluctuations), affecting the
tunneling probability at various temperatures, can be incorporated into a
single formula.

The tunneling current is obtained from the balance equation
\begin{equation}
I=e\sum_{\mathbf{k}_{||}\mathbf{,k}_{||}^{\prime }}W_{\mathbf{k}_{||}\mathbf{%
,k}_{||}^{\prime }}n_{\mathbf{k}_{||}}^{L}(1-n_{\mathbf{k}_{||}^{\prime
}}^{R})-W_{\mathbf{k}_{||}^{\prime },\mathbf{k}_{||}}n_{\mathbf{k}%
_{||}}^{R}(1-n_{\mathbf{k}_{||}^{\prime }}^{L})\,,
\end{equation}
where $W_{\mathbf{k}_{||}\mathbf{,p}_{||}}$ is a transition rate through the
resonant center and $n_{\mathbf{k}_{||}}^{L/R}$ are the distribution
functions for electrons in the left and right layers, respectively. In the
linear response regime, the conductance of a bi-layer tunneling junction is
\begin{equation}
G=-e^{2}\int d\epsilon d\epsilon ^{\prime }W_{\epsilon ,\epsilon ^{\prime }}%
\bigg[\frac{\partial n_{\epsilon }}{\partial \epsilon }(1-n_{\epsilon
^{\prime }})+\frac{\partial n_{\epsilon ^{\prime }}}{\partial \epsilon
^{\prime }}n_{\epsilon }\bigg]\,,  \label{a1}
\end{equation}
where $\epsilon \equiv \epsilon _{\mathbf{k}_{||}},\epsilon ^{\prime }\equiv
\epsilon _{\mathbf{k}_{||}^{\prime }}$ and $\epsilon _{\mathbf{k}_{||}}$ is
the dispersion for the in-plane motion. The transition rate can be expressed
in terms of the resonant tunneling amplitude\cite{Glazman_1988}
\begin{equation}
W_{\epsilon ,\epsilon ^{\prime }}=\lim_{t\rightarrow \infty }\frac{1}{t}%
\langle U_{\epsilon ,\epsilon ^{\prime }}^{\dagger }(t)U_{\epsilon ,\epsilon
^{\prime }}(t)\rangle \,,
\end{equation}
where the transition amplitude is given by%
\begin{widetext}
\begin{eqnarray}
U_{\epsilon,\epsilon'}(t)=-g_{ca} g_{cd}e^{i(\epsilon-\epsilon')t}\int_0^td\tau_1e^{-i(\epsilon_0-\epsilon'-i\Gamma)\tau_1}\int_0^{\tau_1}d\tau_2
e^{i(\epsilon_0-\epsilon-i\Gamma)\tau_2}{\rm Texp}\left\{ -i\int_{\tau_2}^{\tau_1}\hat{\phi}(t)dt\right\}\,.
\end{eqnarray}
Integrating out the Gaussian field $\hat{\phi}$, one gets 
\begin{equation}
W_{_{\epsilon ,\epsilon ^{\prime }}}=g_{ca}^{2}g_{cd}^{2}\int_{-\infty
}^{\infty }dt_{1}e^{-it_{1}(\epsilon -\epsilon ^{\prime })}\int_{0}^{\infty
}dt_{2}dt_{3}e^{t_{2}(-i(\epsilon _{0}-\epsilon )-\Gamma )+t_{3}(i(\epsilon
_{0}-\epsilon ^{\prime })-\Gamma )}V(t_{1},t_{2},t_{3})\,,  \label{bb1}
\end{equation}
where $\Gamma =\Gamma _{L}+\Gamma _{R}$, 
$\Gamma _{L}=\sum_{\mathbf{p}_{||}}g_{ca}^{2}\delta (\epsilon _{0}-\epsilon
_{\mathbf{p}_{||}}),\,\,\Gamma _{R}=\sum_{\mathbf{p}_{||}}g_{cd}^{2}\delta
(\epsilon _{0}-\epsilon _{\mathbf{p}_{||}})$, 
\begin{eqnarray}
&&V(t_{1},t_{2},t_{3})=\exp \bigg\{\frac{1}{\pi }\int \frac{d\omega }{\omega
^{2}}S\left( \omega ,z_{i}\right) \bigg[i\omega (t_{3}-t_{2})-e^{i\omega
t_{3}}-e^{-i\omega t_{2}}-e^{i\omega t_{1}}+e^{i\omega
(t_{1}+t_{3})}+e^{i\omega (t_{1}-t_{2})}-e^{i\omega (t_{3}-t_{2}+t_{1})}
\notag \\
&&-\coth \left( \frac{\omega }{2T}\right) \left( e^{i\omega
t_{3}}+e^{-i\omega t_{2}}-2+e^{i\omega t_{1}}-e^{i\omega
(t_{1}+t_{3})}-e^{i\omega (t_{1}-t_{2})}+e^{i\omega
(t_{1}+t_{3}-t_{2})}\right) \bigg]\bigg\}\,.  \label{v123}
\end{eqnarray}
\end{widetext} Here $S\left( \omega ,z_{i}\right) $ is the local spectral
function of potential fluctuations, as measured at the resonant site
\begin{equation}
S\left( \omega ,z_{i}\right) =\frac{4}{\pi }\int \frac{d^{2}q_{||}}{\left(
2\pi \right) ^{2}}\mathrm{Im}D^{R}(\omega ,\mathbf{q}_{||},z_{i},z_{i}),
\label{spectral}
\end{equation}
where the retarded propagator of $\hat{\phi}$ is \begin{widetext}
\begin{equation}
D^{R}(\omega ,\mathbf{q}_{||},z,z^{\prime })=-i\int_{0}^{\infty }dt\int
d^{2}r_{||}e^{i(\omega t-\mathbf{q}_{||}\cdot \mathbf{r}_{||})}\langle \left[
\hat{\phi}\left( \mathbf{r}_{||},z,t\right) ,\hat{\phi}\left( \mathbf{0}%
,z^{\prime },0\right) \right] \rangle .
\end{equation}
\end{widetext}
As we assume that the metal is translationally invariant
in the in-plane direction but periodic in the $z$-direction, the spectral
function, in general, depends on $z_{i}.$ In the absence of interactions,
Eq. (\ref{bb1}) reproduces the Breight-Wigner formula. For the
electron-phonon interaction, $\hat{\phi}$ is a deformation potential at the
resonant site. In that case Eq.(\ref{bb1}) reproduces the result of Ref.
\onlinecite{Glazman_1988}.

From now on, we consider the case of a resonant-impurity band, assuming that
the resonant centers are randomly distributed over the inter-layer spacing
while their energies are uniformly distributed in the interval $E_{b}$
around the Fermi energy. Assuming that $E_{b}\gg T$ and averaging over $%
z_{i} $ and $\varepsilon _{0}$, we simplify Eq.(\ref{a1}) further  to
\begin{equation}
\!\!\!G\!=\!e^{2}\!\int_{-E_{b}}^{E_{b}}\!\!\!W(\epsilon )\!\bigg[%
\!1\!-\!\!\coth \left( \frac{\epsilon }{2T}\right) \!\!+\!\frac{\epsilon }{2T%
}\frac{1}{\sinh ^{2}\left( \frac{\epsilon }{2T}\right) }\!\bigg](d\epsilon
)\,,  \label{bb2}
\end{equation}
where \begin{widetext}
\begin{equation}
W(\epsilon )=W_{0}\int_{-\infty }^{\infty }dt\exp \bigg\{-i\epsilon
t+\int_{0}^{\infty }\frac{d\omega }{\omega ^{2}}S_{m}(\omega )\bigg[(1-\cos
(\omega t))\coth \left( \frac{\omega }{2T}\right) -i\sin (\omega t)\bigg]%
\bigg\}\,.  \label{cc6}
\end{equation}
\end{widetext} Here $W_{0}$ is the resonant transition probability in the
absence of interaction and $S_{m}(\omega )=S\left( \omega ,d/2\right) $ is
the spectral function at the position of the most efficient resonant center,
i.e., in the middle of the spacing between the layers. The local spectral
function $S_{m}(\omega )$ contains all the information about the fluctuating
field.

Eqs.(\ref{bb2},\ref{cc6}) describe the average conductance for a single
resonant impurity. Since a layered metal can be viewed as sequence of
tunneling junctions connected in series, the resonant tunneling conductivity
of the whole crystal is related to the bi-layer conductance by the Ohm's
law: $\sigma _{\mathrm{res}}=Gd.$ Performing integration over energy in Eq.(%
\ref{bb2}), we arrive at the general result for the boson-assisted tunneling
conductivity: \begin{widetext}
\begin{equation}
\!\!\!\!\!\!\!\sigma _{\mathrm{res}}\!=\!\sigma _{\mathrm{el}%
}\!\int_{-\infty }^{\infty }\!\!\!\!dt\frac{i\pi T^{2}t}{\sinh ^{2}\left(
\pi Tt+i0\right) }\exp \bigg\{\int_{0}^{\infty }\frac{d\omega }{\omega ^{2}}%
S_{m}(\omega )\bigg[(1-\cos (\omega t))\coth \left( \frac{\omega }{2T}%
\right) -i\sin (\omega t)\bigg]\bigg\}\,, \label{a5}
\end{equation}
\end{widetext} where $\nolinebreak {\sigma _{\mathrm{el}}\simeq \pi
e^{2}\Gamma n_{\mathrm{imp}}a_{0}d/E_{b}}$\cite{Larkin_Matveev} is a
resonant conductivity of free electrons, $n_{\mathrm{imp}}$ is the number of resonant impurities per unit
volume,
and $a_{0}$ is the localization
radius of a resonant state.

We note a similarity between our result (\ref{a5}) and those of two other
problems: the zero bias anomaly  in the tunneling current \cite{AA-review}
and ``dissipative localization'' of a particle coupled to the thermal bath
[the Caldeira-Leggett (CL) model \cite{CL}]. Indeed, all of these problems
are related and describe different variations of the M\"{o}ssbauer effect
\cite{Lipkin}. In the case of tunneling through a resonant impurity, its
occupied and empty states play the role of two pseudo-spin states of the CL
particle, whereas the fluctuating electrostatic potential is analogous to
the thermal bath of harmonic oscillators. While this similarity is not
formally obvious for tunneling through a single resonant center, it becomes clear
after averaging over the ensemble of levels. Despite the similarity, there
are also some difference with the CL model.
In the CL model, the effect of the environment is incorporated
phenomenologically, via the parameters of noise correlation function. A more
microscopic resonant-tunneling model treats explicitly fluctuations of the
electrostatic potential, incorporated in the spectral function $S_{m}(\omega
)$. As the temperature changes, so does the characteristic frequency scale
and, consequently, the most efficient bosonic mode. This results in a number
of intermediate regimes of assisted tunneling, each of them
corresponding to its own effective CL model.

The next two Sections are devoted to a quantitative analysis of some of the most
common sources of interaction: the electron-phonon and electron-electron
ones.

\subsection{Phonon-assisted tunneling}
\label{sec:eph}

We assume that phonons are isotropic and that the electron-phonon
interaction is of the deformation-potential type. Although phonon modes in
real layered metals are anisotropic, this anisotropy is still much weaker
than the anisotropy of electron spectra. Anisotropy of phonon modes can be
incorporated into the theory without any difficulties. The retarded
correlation function of the deformation potential is given by
\begin{equation}
D^{R}(\omega ,q)=\alpha _{\mathbf{q}}^{2}\frac{\omega _{q}}{(\omega
+i0)^{2}-\omega _{q}^{2}}\,,  \label{dp}
\end{equation}
where ${\alpha _{\mathbf{q}}^{2}\equiv \Lambda ^{2}}q{^{2}/\rho \omega _{%
\mathbf{q}},}$ $\Lambda $ is the deformation potential constant,
and $\rho $ is the atomic mass density. In the elastic continuum
model, the spectral function is translationally invariant in the
$z$-direction. For acoustic phonons $\omega
_{\mathbf{q}}=sq\theta \left( q_{D}-q\right) $ , Eq.(\ref{spectral},\ref{dp}%
)) gives
\begin{equation}
S_{m}^{\mathrm{e-ph}}(\omega )=-\lambda \theta \left( \omega _{D}-\omega
\right) \frac{\omega ^{3}}{\omega _{D}^{2}}\,\,,  \label{dd4}
\end{equation}
where
\begin{equation}
\lambda \equiv \Lambda ^{2}\omega _{D}^{2}/\rho s^{5}\pi ^{2}  \label{lambda}
\end{equation}
is the dimensionless coupling constant for localized electrons,
$\omega _{D}=sq_{D},$ and $\theta \left( x\right) $ is the
step-function. The interaction with acoustic phonons
corresponds to the super-Ohmic regime of the CL model. Using Eqs. (\ref{a5},%
\ref{dd4}), we obtain \begin{widetext}

\begin{eqnarray}\label{cc5}
&&\!\!\!\!\!\!\!\sigma _{\mathrm{res}}\!=\!\sigma _{\mathrm{el}%
}\!\int_{-\infty }^{\infty }\!\!\!\!dt\frac{i\pi T^{2}t}{\sinh ^{2}\left(
\pi Tt+i0\right) }e^{-\!\lambda f(t)},\,\mathrm{where}  \label{res_tun_a} \\
&&\hspace{-1cm}f(t)\!=\!\int_{0}^{\omega _{D}}\!\!\!d\omega \frac{\omega }{%
\omega _{D}^{2}}\bigg[\!\left( (1\!-\!\cos (\omega t)\right)\left[\coth
\left( \!\frac{\omega}{2T}\!\right)-1\right] \!+\left(1-e^{i\omega
t}\right)\bigg].\label{sigmares_b}
\end{eqnarray}
\end{widetext}
In the absence of the electron-phonon interaction ($\lambda =0$), $\sigma _{%
\mathrm{res}}=\sigma _{\mathrm{el}}$ .%

Notice that the electron-phonon interaction is much stronger for
localized electrons than for thewe  band ones. Indeed, the
dimensionless coupling constant for bulk electrons which
determines, e.g., the mass-renormalization, is of order of unity
for most metals: $\zeta =\alpha _{q_{D}}^{2}\left( q_{D}\right)
q_{D}/v_{F}s\sim 1$. The coupling constant between localized
electrons and phonons is larger than $\zeta $ by at least the
Migdal parameter: $\lambda \sim \zeta \left( k_{F}d\right)
(v_{F}/s)\gg 1.$ Therefore, one needs to consider a
non-perturbative regime of phonon-assisted tunneling. Analyzing
Eq.(\ref{cc5})(for details, see Appendix \ref{phonon_assisted}),
one finds that resonant tunneling is exponentially suppressed at
$T=0$:
\begin{equation}
\sigma _{\text{res}}(T=0)=\sigma _{\text{el}}e^{-\lambda /2}.
\end{equation}
At finite $T$, we find 
\begin{equation}
\frac{\sigma _{\text{res}}}{\sigma _{\text{el}}}=\left\{
\begin{array}{l}
e^{-\lambda /2}\left( 1+\frac{\pi ^{2}\lambda }{3}\left( \frac{T}{\omega _{D}%
}\right) ^{2}\right) \,,\,\,\,T\ll \frac{\omega _{D}}{\sqrt{\lambda }}\,\,,
\\
\exp \left( -\frac{\lambda }{2}+\frac{\lambda }{3}\left( \frac{\pi T}{\omega
_{D}}\right) ^{2}\right) \,,\,\frac{\omega _{D}}{\sqrt{\lambda }}\ll \!T\ll
\omega _{D} \\
\sqrt{\frac{3\pi ^{3}T}{4\lambda \omega _{D}}}\exp \left( -\frac{\lambda
\omega _{D}}{12T}\right) \,,\,\,\,\omega _{D}\ll T\ll \lambda \omega
_{D}\,\,, \\
1-\frac{\lambda }{9}\frac{\omega _{D}}{T}\,\,,\,\,\,T\gg \lambda \omega _{D}.
\end{array}
\right.   \label{cc7}
\end{equation}
The low-temperature limit ($T\ll \omega _{D}/\sqrt{\lambda }$), reproduces
the known result of Ref.~[\onlinecite{Glazman_1988}]. As we see, $\sigma _{\text{res }%
}$ increases with $T,$ resembling the zero-bias anomaly in disordered metals
and M\"{o}ssbauer effect. 
At high temperatures ($T\gg T_{s}= \lambda \omega _{D}$), $\sigma _{%
\text{res}}$ saturates at the non-interacting value ($\sigma _{\text{el}}$),
because thermal activation cannot make the conductivity larger than in the
absence of phonons. Notice that, in contrast to the phenomenological model
of Ref.~[\onlinecite{levin}], there is no simple relation between the $T$-dependences
of $\sigma _{c}^{B}$ and $\sigma _{\mathrm{res}}$.

The temperature dependence of the tunneling conductivity also
resembles the polaronic behavior
\cite{lang_firsov,polaron_mckenzie,polaron_schofield}. Indeed,
phonon-assisted tunneling can be interpreted in terms of a local
polaron formation. However, we emphasize that polarons are stable
here because the velocity of on-site electrons is determined by
the tunneling width and thus very small. In the absence of
resonant impurities, a metal with a large Migdal parameter
$v_{F}/s\gg 1$ cannot sustain stable polarons, as they emit
phonons and decay.

\subsection{Electron-assisted tunneling}
\label{sec:ee} The effect of the interaction between on-site and
conduction electrons on resonant tunneling is qualitatively
similar to the effect of phonons, as a tunneling electron drags a
surrounding cloud of electrons. This mechanism is also known as
the ``electronic polaron'' \cite{Mahan}.

\begin{figure}[h]
\label{fig4}\vspace{1cm} 
\includegraphics[angle=0,width
=0.5\textwidth] {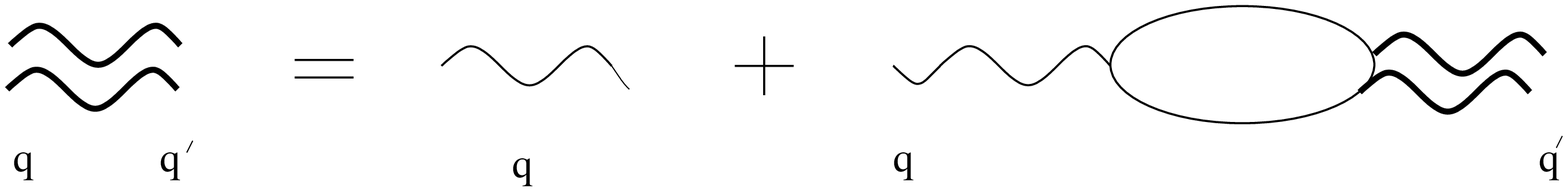} 
\caption{RPA series for electrons on a lattice. The incoming and
outgoing bosonic momenta can differ by an arbitrary reciprocal
lattice vector.} \label{fig:rpa}
\end{figure}

The on-site Coulomb potential is produced by all mobile electrons in a
metal. The Green's function of electrons on a lattice is not translationally
invariant. In the tight-binding model with spectrum (\ref{spectrum}), it is
given by
\begin{equation*}
G^{R}\left( \varepsilon ,\mathbf{k}_{||},k_{z},k_{z}^{\prime }\right)
=\sum_{b}\delta _{k_{z},k_{z}^{\prime }+b}G_{0}^{R}\left( \varepsilon ,%
\mathbf{k}\right) ,
\end{equation*}
where $G_{0}^{R}\left( \varepsilon ,\mathbf{k}\right) =\left( i\varepsilon
-\epsilon _{\mathbf{k}}+i/2\tau \right) ^{-1}$ , $b=2\pi n/d$ is the
reciprocal lattice vector, and $\mathbf{k=}\left( \mathbf{k}%
_{||},k_{z}\right) .$ The screened Coulomb potential is given by
the RPA series (Fig. \ref{fig:rpa}). Because the in-plane motion
is free, the in-plane momentum is conserved at the vertices.
However, it is the out-of-plane quasi-momentum rather than the
momentum that is conserved at the vertices. In addition to normal
scattering, where the incoming and out-going bosonic momenta of
the polarization bubble are the same, the series in Fig.\ref
{fig:rpa} also accounts for Umklapp processes, in which $q_{z}$ and $%
q_{z^{\prime }}$ differ by $b$.

Summing up the geometric series, we get for the Fourier transform of the
dynamic screened potential \begin{widetext}
\begin{equation}
D^{R}(\omega ,\mathbf{q}_{||},q_{z},q_{z}^{\prime })=D_{0}(\mathbf{q}%
_{||},q_{z})\delta (q_{z}-q_{z}^{\prime })-\frac{D_{0}(\mathbf{q}%
_{||},q_{z})\Pi ^{R}(\omega ,\mathbf{q})\sum_{b}D_{0}(\mathbf{q}%
_{||},q_{z}+b)\delta (q_{z}^{\prime }-q_{z}-b)}{1+\sum_{b}D_{0}(\mathbf{q}%
_{||},q_{z}+b)\Pi ^{R}(\omega ,\mathbf{q})}.  \label{rpa1}
\end{equation}
\end{widetext} Here,
\begin{equation}
D_{0}(\mathbf{q}_{||},q_{z})=\frac{4\pi e^{2}}{q_{\parallel }^{2}+q_{z}^{2}}
\end{equation}
is the bare Coulomb potential and $\Pi ^{R}(\omega ,\mathbf{q})$ is the
analytic continuation of the Matsubara polarization bubble
\begin{equation}
\Pi (i\omega ,\mathbf{q}_{||},q_{z})=T\sum_{\varepsilon }\int \frac{d^{3}k}{%
\left( 2\pi \right) ^{3}}G_{0}\left( i\varepsilon +i\omega ,\mathbf{k+q}%
\right) G_{0}\left( i\varepsilon ,\mathbf{k}\right) ,  \label{bubble}
\end{equation}
where $G_{0}\left( i\varepsilon ,\mathbf{k}\right) =\left( i\varepsilon
-\epsilon _{\mathbf{k}}+i\mathrm{sgn}\varepsilon /2\tau \right) ^{-1}$ and $%
\mathbf{q=}\left( \mathbf{q}_{||},q_{z}\right) .$ The potential at the
position of the most efficient resonant center, $z_{i}=d/2$, is obtained
from Eq.(\ref{rpa1}) as \begin{widetext}
\begin{equation*}
D^{R}\left( \omega ,\mathbf{q}_{||},d/2,d/2\right) =\int \frac{dq_{z}}{2\pi }%
\int \frac{dq_{z}^{\prime }}{2\pi }e^{i(q_{z}-q_{z}^{\prime
})d/2}D^{R}\left( \omega ,\mathbf{q}_{||},q_{z},q_{z^{\prime }}\right) =%
\frac{2\pi e^{2}}{q_{||}}\!+\!\int \frac{dq_{z}}{2\pi }\frac{D_{0}\Pi
^{R}A_{1}}{1+\Pi ^{R}A_{2}},
\end{equation*}
\end{widetext} 
where $A_{1}$ and $A_{2}$ are the lattice sums which can be performed
explicitly \begin{widetext}
\begin{eqnarray}
A_{1} &=&\sum_{b}D_{0}(\mathbf{q}_{||},q_{z}+b)=\frac{2\pi e^{2}d}{%
q_{\parallel }}\frac{\sinh (q_{\parallel }d)}{\cosh (q_{\parallel }d)-\cos
(q_{z}d)} \\
A_{2} &=&\sum_{b}D_{0}(\mathbf{q}_{||},q_{z}+b)e^{-ibd/2}=\frac{4\pi e^{2}d}{%
q_{\parallel }}\frac{\cos (q_{z}d/2)\sinh (q_{\parallel }d/2)}{\cosh
(q_{\parallel }d)-\cos (q_{z}d)}.
\end{eqnarray}
\end{widetext} In a strongly layered metal, the polarization operator (\ref
{bubble}) depends on $q_{z}$ only weakly, because the interlayer hopping $%
J_{c}$ is small and the electron dispersion is almost two-dimensional. If
this dependence can be neglected completely, the integration over $q_{z}$
can be readily performed with the result
\begin{equation}
D^{R}\left( \omega ,\mathbf{q}_{||},d/2,d/2\right) =\frac{2\pi e^{2}}{%
q_{\parallel }}\tanh \left( \frac{q_{\parallel }d}{2}\right) \coth \left(
\frac{kd}{2}\right) ,  \label{dd1}
\end{equation}
where (complex) momentum $k$ is defined by the following equation:
\begin{equation}
\cosh (kd)=\cosh (q_{\parallel }d)+\frac{2\pi e^{2}}{q_{\parallel }}\Pi
(\omega ,q_{\parallel })\sinh (q_{\parallel }d)\,.
\end{equation}
If the effective range of the screened Coulomb potential is much larger than
the lattice spacing, Umklapp scattering is strongly suppressed.
Consequently, in the limit of $\kappa _{3}d\ll 1,$ where $\kappa
_{3}^{2}=4\pi e^{2}\nu _{3}$ is the (square of) screening wave vector, and $%
\nu _{3}$ is the three-dimensional density of states, Eq.(\ref{dd1}) reduces
to the continuum limit result. For finite values of $\kappa _{3}d,$ the role
of Umklapp processes is quite important.

Eq.(\ref{dd1}) can be reproduced in an alternative way. In the limit of $%
J_{c}\rightarrow 0$, the motion of electrons between different layers is
forbidden. Therefore, the problem is equivalent to the one of screening of
an external charge by parallel conducting layers\cite{Mishchenko_2001}.
Although both approaches are equivalent for $J_{c}=0$, the RPA method is
more general since, unlike electrostatics, it allows one to consider the
case of finite $J_{c}$ as well.

The polarization bubble can be calculated explicitly in the diffusive ($%
\omega \tau \ll 1$)
\begin{equation}
\Pi ^{R}(\omega ,q)=\nu _{3}\frac{D_{\parallel }q_{\parallel }^{2}+8\tau
J_{c}^{2}\sin ^{2}(q_{z}d/2)}{D_{\parallel }q_{\parallel }^{2}+8\tau
J_{c}^{2}\sin ^{2}(q_{z}d/2)-i\omega }\,,  \label{pi_diff}
\end{equation}
and ballistic ($\omega \gg \tau ^{-1}$) limits
\begin{equation}
\Pi ^{R}(\omega ,q)\!=\!\nu _{3}\bigg[1+\frac{i\omega }{v_Fq_{\parallel }}%
\bigg]\,,\mathrm{for}\,\,v_{F}q_{\parallel }\gg \min \left\{
J_{c}dq_{z},\omega \right\} \,.  \label{pi_ball}
\end{equation}
Here $D_{||}=v_{F}^{2}\tau/2 $ is the in-plane diffusion coefficient. The
additional assumption of large $q_{||},$ employed in Eq.(\ref{pi_ball}),
will be justified later.

The local spectral weight $S_{m}^{\mathrm{e-e}}\left( \omega \right) $ for
the electron-electron interaction has two distinct forms of the $\omega $
dependence. Assuming that the motion of electrons is ballistic and strictly
two-dimensional, we expand the imaginary part of the potential in Eq.(\ref
{dd1}) to first order in $\omega /v_{F}q_{||}$ and perform the integration
over $q_{||}.$ Typical value of $q_{||}$ turns out to be large: of order of
the 2D screening wave vector $\kappa _{2}=2\pi e^{2}\nu _{2},$ where $\nu
_{2} $ is the 2D density of states. On the other hand, typical value of $%
\omega $ are determined by the temperature. Therefore, the assumption of $%
\omega /v_{F}q_{||}$ is satisfied for all reasonable temperatures: $T\ll
v_{F}\kappa _{2}.$ As the result comes from the first-order term in $\omega
/v_{F}q_{||},$ the resulting spectral density is linear in $\omega $%
\begin{equation}
S_{m}^{\mathrm{e-e}}(\omega )=-\eta _{B}\omega ,  \label{ohmic}
\end{equation}
where $\eta _{B}$ is the effective coupling for the electron-electron
interaction
\begin{equation}
\eta _{B}=\frac{e^{2}}{v_{F}}g_{B}\left( \kappa _{2}d\right)  \label{eta_b}
\end{equation}
and \begin{widetext}
\begin{equation}
g_{B}\left( x\right) =x\int_{0}^{\infty }dy\frac{\coth \frac{y}{2}}{\cosh y}%
\frac{1}{\left( y+x\tanh \frac{y}{2}\right) ^{1/2}\left( y+x\coth \frac{y}{2}%
\right) ^{3/2}}\approx \left\{
\begin{array}{l}
1\,,\,\,\,\,\,\text{for }x\ll 1 \\
\frac{\pi }{2x}\,,\,\,\,\,\,\text{for }x\gg 1.
\end{array}
\right. .  \label{eta_b_vs_d}
\end{equation}
\end{widetext} Large in-plane momentum transfers also help to justify the
assumption of two-dimensionality. Indeed, when calculating the polarization
operator, one has to compare the difference of dispersions for the 2D and $c$%
-axis motions, i.e., $\delta \epsilon _{||}=\epsilon _{\mathbf{k}_{||}+%
\mathbf{q}_{||}}-\epsilon _{\mathbf{k}_{||}}\sim v_{F}q_{||}\sim v_{F}\kappa
_{2}$ and $\delta \epsilon _{z}=\epsilon _{k_{z}+q_{z}}-\epsilon
_{q_{z}}\sim Jdq_{z}.$ Typical values of $q_{z}$ are of order $\min \left\{
\kappa _{2},d^{-1}\right\} .$ Hence, neglecting the $z$-component of the
electron dispersion is justified as long as $J\ll v_{F}\kappa _{2}/d\min
\left\{ \kappa _{2},d^{-1}\right\} ,$ which is the case for any real layered
metal. In the CL terminology \cite{CL}, Eq.(\ref{ohmic}) corresponds to
Ohmic regime. Substituting Eq.(\ref{ohmic}) into Eq.(\ref{a5}) and cutting
off the ultraviolet logarithmic singularity in the $\omega $ integral at
some high-energy scale $E_{0}$ , we evaluate the integral for large $E_{0}t$
as \begin{widetext}
\begin{equation}
\int_{0}^{E_{0}}\frac{d\omega }{\omega }\bigg[(1-\cos (\omega t))\coth
\left( \frac{\omega }{2T}\right) -i\sin (\omega t)\bigg]=\ln (E_{0}t)-\frac{i\pi}{2}\rm{sgn}(t)+\gamma
+\dots ,
\end{equation}
\end{widetext} where $\gamma =0.577...\,$ is the Euler's constant. Rescaling $t$
by $T^{-1}$ in the remaining integral of Eq.(\ref{a5}), we find
that the Ohmic regime of the spectral function corresponds to a
power-law scaling of the resonant conductivity
\begin{equation}
\sigma _{\mathrm{res}}(T)/\sigma_{\rm el}=C\left( \eta _{B}\right)\left( \frac{T}{E_{0}}\right) ^{\eta _{B}},\,\,\,  \label{sigma_b}
\end{equation}
where the regularization dependent prefactor $C\left(x\right)$
(for a hard cutoff regularization of the frequency integral) is
given by
\begin{equation}
C(x)=\cos\left(\frac{\pi x}{2}\right)e^{-\gamma x}\,.
\end{equation}

As the distance between the planes increases, the tunneling exponent $\eta _{B}$
[cf. Eqs.(\ref{eta_b},\ref{eta_b_vs_d})] decreases from its nominal value,
given by the dimensionless electron-electron coupling constant in a bulk
metal, to zero, which is to be expected.

The expansion of the polarization bubble in $\omega /v_{F}q_{||}$, which led
us to Eq.(\ref{sigma_b}) works as long as the subsequent integral over $%
q_{||}$ converges (and typical values of $q_{||}$ are determined
by the ultraviolet parameters of the theory). This is the case for
the leading term in the expansion; however, the next order term
diverges logarithmically in the
infrared. In the ballistic limit, the divergence results in a subleading $%
\omega ^{2}\ln \left| \omega \right| $ correction to the spectral function,
which is not relevant at low temperatures. However, one of the factors of $\omega $ is replaced by $%
\tau ^{-1}$ in the diffusive limit ($%
\omega \ll \tau ^{-1})$,  and the subleading term becomes larger
than the leading, linear-in-$\omega $ term. The resulting $\omega
\ln \left| \omega \right| $ behavior of the spectral function can
be obtained accurately by starting with the diffusive rather than
ballistic form of the polarization bubble
[Eq.(\ref{pi_diff})]. In the diffusive regime, one can still neglect the $%
q_{z}$-dependent terms in $\Pi ^{R}\left( \omega ,q\right) ,$ which amounts
to an assumption of purely 2D diffusion, as long as $\nolinebreak {\omega
\gg \omega }_{1}$, where
\begin{equation}
\omega _{1}=J^{2}\tau \kappa _{3}d\left\{
\begin{array}{l}
\kappa _{3}d\,,\,\,\,\,\,\kappa_{3}d\ll 1 \\
1\,,\,\,\,\,\,\kappa _{2}d\gg 1.
\end{array}
\right.   \label{omega_1}
\end{equation}
The resulting spectral weight is a sublinear function of the
frequency
\begin{equation}
S_{m}^{\text{e-e}}(\omega )=\eta _{\mathrm{D}}\omega \log \left( \frac{%
\omega }{\omega _{0}}\right) ,  \label{superohmic}
\end{equation}
where
\begin{equation}
\eta _{\mathrm{D}}=\frac{1}{4\pi \nu _{2}D_{||}}\left\{
\begin{array}{l}
\kappa _{2}d/2\,,\,\,\,\,\,\kappa _{2}d\ll 1 \\
1\,,\,\,\,\,\,\kappa _{2}d\gg 1,
\end{array}
\right.
\end{equation}
\begin{equation}
\omega _{0}=\left\{
\begin{array}{l}
D_{||}\kappa _{2}^{2}\,,\,\,\,\,\,\kappa _{2}d\ll 1 \\
\kappa _{2}dD_{||}/l^{2},\,\,\,\,\,\kappa _{2}d\gg 1,
\end{array}
\right.
\end{equation}
and $l=v_{F}\tau $ is the mean free path.
The corresponding conductivity
\begin{equation}
\sigma _{\mathrm{res}}(T)=\sigma _{\mathrm{el}}\exp \bigg\{\eta _{2\mathrm{D}%
}\log \left( T\tau \right) \log \left( \frac{\omega _{0}^{2}\tau }{T}\right) %
\bigg\}\,  \label{sigma_d}
\end{equation}
increases with temperature faster than any power-law. Interestingly, the
temperature dependence of the resonant conductivity is similar to that of
the zero-bias anomaly (ZBA) in the 2D diffusive case\cite{AA-review,Finkelstein83}.
The difference between the two results is in the
dimensionless parameters: the dimensionless conductance, $g=\nu _{2}D_{||},$
that controls ZBA, is replaced by $\eta _{\text{2D}}$ for the resonant
tunneling case.

For frequencies smaller than $\omega _{1}$, cf. Eq.[\ref{omega_1}], typical  $q_{||}$'s are large, which means the diffusion approximation breaks
down again and the ballistic one should be used instead. We thus conclude
that the resonant tunneling conductivity is given by the diffusive limit
[Eq.(\ref{sigma_d})] for temperatures in the interval $\omega _{1}\ll T\ll
1/\tau $ and by the ballistic limit [Eq.(\ref{sigma_b})] for the rest of the
temperatures.

\section{Discussion of the results}
\label{sec:discussion}
Several comments are in order.

i) Formally speaking, the power-law scaling of the electron-assisted
tunneling conductivity in the ballistic regime [Eq.(\ref{sigma_b})]
saturates for $T\gtrsim E_{0}$. The value at saturation is the same as for
the electron-phonon case: $\sigma _{\mathrm{el}}.$ However, $E_{0}$ is of
order of the plasma frequency, so that the electron-assisted mechanism of
tunneling leads to a growth of conductivity for all reasonable temperatures.
This feature may be used in experiment to separate the electron- and
phonon-assisted mechanism: because the saturation temperature for the phonon
mechanism may be not too high, the growth of the total conductivity up to
the highest temperatures is indicative of the electron-assisted mechanism.

ii) In contrast to the case of the interaction corrections to
tunneling and transport conductivities in the ballistic regime
\cite{rudin,zna}, which are determined by the interaction on a
large spatial scale (of order of the ballistic Thouless length
$v_{F}/T$), the electron-assisted tunneling conductivity is
determined by the interaction at small distances (of order of the
screening radius). At these distances, the RPA works only for weak
interactions, i.e., for $e^{2}/v_{F}\ll 1,$ and Eq.(\ref{sigma_b})
is valid, strictly speaking, only in the perturbative regime. It
is reasonable to assume, however, that if the interaction is not
weak, the tunneling exponent in Eq.(\ref{sigma_b}) is replaced by
a non-universal quantity of order unity.

iii) In a real system, both assisted mechanisms--the electron-phonon and
electron-electron ones-operate simultaneously. As we have explained before,
the electron-phonon interaction is strong while the electron-electron one
must be considered to be weak. To the extent that one can neglect the mutual
influence of these two interactions, the total spectral function in Eq.(\ref
{a5}) is a sum of two contributions $S_{m}\left( \omega \right) =S_{m}^{%
\mathrm{e-ph}}+S_{m}^{\mathrm{e-e}}(\omega ).$ Since $\eta _{B}\ll 1,$ one
can expand the total conductivity $\sigma _{\mathrm{res}}^{\mathrm{tot}}$ to
first order in $S_{m}^{\mathrm{e-e}}(\omega ),$ which yields
\begin{equation}
\sigma _{\mathrm{res}}^{\mathrm{tot}}=\sigma _{\mathrm{res}}\left( 1+\eta
_{B}\ln \frac{E_{0}}{T}\right) ,  \label{total}
\end{equation}
where $\sigma _{\mathrm{res}}$ is the electron-phonon contribution given by
Eq.(\ref{cc7}). Ignoring the logarithmic temperature dependence and using
the low-temperature result for $\sigma _{\mathrm{res}}$ [first line in Eq.(%
\ref{cc7})], one sees that the electron-electron contribution
dominates over the electron phonon-one for temperatures below a
characteristic temperature
\begin{equation}
T_{L}=\sqrt{\eta _{B}/\lambda }\omega _{D}\,.  \label{tl}
\end{equation}
Using Eq.(\ref{lambda}) for $\lambda ,$ estimating $\eta _{B}$ as $%
e^{2}/v_{F}$ and also the deformation-potential constant as $\Lambda \sim
Ms^{2}\sim e^{2}/a_{0},$ where $M$ is the atomic mass and $a_{0}$ is the
lattice constant, we obtain $T_{L}\sim \sqrt{s/v_{F}}\omega _{D}\sim 10$ K.
For temperatures above $T_{L}$ but below the saturation temperature for the
electron-phonon mechanism $T_{s}=\lambda \omega _{D},$ the electron-phonon
mechanism dominates. For $T\gtrsim T_{s}$ the electron-electron interaction
wins over again. Once the electron-electron mechanism becomes the dominant
one, the logarithms following the lowest order one in Eq.(\ref{total}) can
be summed up into a power law. Therefore, a signature for the
electron-electron mechanism is a power-law scaling of the conductivity (with
an exponent of order unity) both at low and high temperatures, combined with
an absence of saturation at high temperatures, as it was explained in item
i) of this Section. The power-law increase of the conductivity with an
exponent close to one has been observed in graphite both at low and high
temperatures \cite{hebard_unpub}.

iv) Another mechanism, competing with those considered in this paper, is the
zero-bias anomaly (ZBA), i.e., the interaction corrections to the tunneling
density of states. This mechanism leads to the temperature dependence of the
resonant tunneling conductivity via the temperature dependence of the
tunneling widths even in the absence of assisted processes. The interaction
correction to the tunneling density of states for a 2D electron system can
be divided into parts. The first contribution, $\delta \nu _{\text{cl}%
}\left( \varepsilon \right) ,$ is determined by the interaction of
electrons in the absence of impurities
\cite{reizer,Mishchenko_2001}. In a layered metal, $\delta \nu
_{\text{cl}}\left( \varepsilon \right) $ is energy-independent for
$\varepsilon $ below the plasmon gap \cite {Mishchenko_2001},
which can be safely assumed to be the case, so $\delta
\nu _{\text{cl}}\left( \varepsilon \right) $ does not contribute to the $T$%
-dependence of the tunneling conductivity. The second
contribution, $\delta \nu _{d}\left( \varepsilon \right) ,$ comes
from the interplay of electron-electron and electron impurity
scatterings. In the perturbation theory, $\delta \nu _{d}\left(
\varepsilon \right) $ is proportional to the inverse dimensionless
conductance, $\left( E_{F}\tau \right) ^{-1}$ times the
logarithmic function of $\varepsilon $ \cite
{AA-review,rudin,Mishchenko_2001} both in the ballistic and
diffusive limits. On the other hand, the electron-assisted
tunneling conductivity in the ballistic regime
[Eq.(\ref{sigma_b})] allows for an expansion in $\eta _{B}\ln
E_{0}/T.$ Therefore, the electron-assisted mechanism wins over the
ZBA one if $\eta _{B}$ is not too small: $\eta _{B}\gg \left(
E_{F}\tau \right) ^{-1}.$ In the diffusive regime, the
electron-assisted and ZBA mechanisms are, generally speaking, of
the same order.

v) In Ref.~[\onlinecite{abrikosov_res}], the insulating-like temperature dependence of $%
\rho _{c}\left( T\right) $ in high-$T_{c}$ cuprates was ascribed
to resonant tunneling through a single resonant level, at energy
$\Delta \varepsilon $ away the Fermi level. No assisted tunneling
mechanisms were invoked: the temperature dependence was coming
entirely from the thermal distribution of conduction electrons. In
this model, the $T$ dependence of the tunneling conductivity is
insulating-like for $T\lesssim \Delta \varepsilon $ and
metallic-like for $T\gtrsim \Delta \varepsilon .$ However,
averaging over the energy levels eliminates the $T$-dependence. We
do not think that an assumption of a single energy level is a very
realistic one and therefore involve boson-assisted mechanisms, in
which the $T$-dependence survives even after averaging over the
energy levels (see, however, Sec. \ref{comparison}).

\section{Comparison with experiment}
\label{comparison} 
\begin{figure}[tbp]
\includegraphics[angle=0,width =0.5\textwidth]{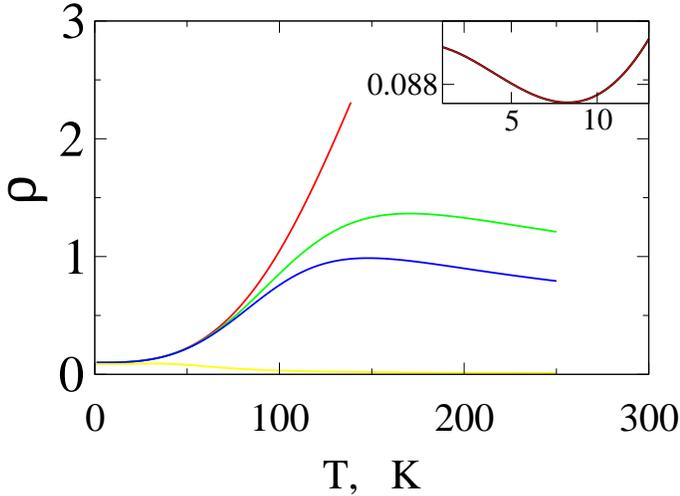}
\caption{resistivity in arbitrary units vs. temperature for various values
of resonant conductivity ($\protect\sigma_{\mathrm{el}}=0,\, 0.00175, \,
0.003,\, 0.2$). On the onset a non-monotonic low temperatures dependence for
$\protect\sigma_{\mathrm{el}}=0.2$.}
\label{fig_all}
\end{figure}

If the total conductivity of a layered metal is a sum of
the band and resonant-tunneling contributions, as specified in Eq.(\ref{cond}%
), then, depending on the parameters of the Boltzmann and
tunneling parts of the conductivity, the total resistivity may
exhibit a variety of $T$ dependences: purely metallic (for weak
tunneling), purely insulating (for strong tunneling), minimum at
low $T$, maximum at high $T$, and both minimum and maximum. Figure
\ref{fig_all} shows some of these behavior for typical band and
resonant conductivities. As an example, we consider a model with a
band resistivity $\rho_B(T)=\rho_0+aT^3$, with $\rho_0=100$ and
$a=10^{-3}$ in arbitrary units. The parameters
for the phonon-assisted resonant conductivity: $\lambda=10$, $\omega_D=200K$. We plot a
total resistivity for different values of the elastic resonant
conductivity, $\sigma_{\rm{el}}$.

The appearance of the maximum in $\rho _{c}\left( T\right) $ has
already been explained in Sec. \ref{sec:intro}: one naturally has
a minimum in the conductivity (and a maximum in resistivity) when
adding up increasing and decreasing functions of the temperature.
The minimum in $\rho _{c}(T)$ arises if $\sigma _{\text{res}}$ is
larger than $\sigma _{B}$ at low temperatures. If quantum
interference effects can be ignored, the low-$T$ dependence of
$\sigma _{B}$ is due to electron-electron interactions. In a Fermi
liquid, $\sigma _{B}=\sigma _{i}\left( 1-a_{\text{U}}T^{2}\tau
/E_{F}\right) ,$ where $\sigma _{i}$ is the residual conductivity
due to impurities and the dimensionless coefficient $a_{\text{U}}$
parameterizes the strength of Umklapp scattering. If the
resonant-tunneling conductivity decreases slower than $T^2$, the
net resistivity shows an insulating behavior. In the
electron-assisted mechanism, the insulating upturn must occur at
low enough temperatures, if exponent $\eta _{B}$ of the
electron-assisted mechanism [Eq.(\ref{sigma_b})] is less than two.
The electron-phonon interaction is a marginal case because the
low-temperature exponent equals precisely to two [see
Eq.(\ref{a1}), first line]. In this case, whether the insulating
upturn occurs or not depends on the magnitude of the
electron-phonon and Umklapp interaction, and also on the amount of
disorder in the sample.

Experiment shows a variety of behaviors in $\rho _{c}(T)$. For example, $%
\rho _{c}$ i) is purely metallic in over doped cuprates; ii) has a minimum
in under doped ones \cite{ginsberg}; iii) is purely insulating in TaS$_{2}$
\cite{frindt}; iv) has both a minimum and a maximum in graphite \cite
{hebard_unpub}; and v) has a maximum in Sr$_{2}$RuO$_{4}$ \cite{srruo}, $%
\kappa $-(BEDT-TTF)$_{2}$-Cu(SCN)$_{2}$ \cite{Analytis},
(Bi$_{0.5}$Pb$_{0.5} $)$_{2}$Ba$_{3}$Co$_{2}$O$_{y}$ and
NaCo$_{2}$O$_{4}$ \cite{Valla}. In our opinion, the most
remarkable behavior is the one with a maximum in $\rho _{c}\left(
T\right) .$ Whereas the insulating upturns can, in principle, be
ascribed to phase transitions, which open gaps over the parts of
the Fermi surface, the metallic behavior of $\rho _{c}\left(
T\right) $ at low temperatures shows unambiguously that we are
dealing with a well-defined metallic state. In all cases cited
above, the insulating behavior at higher temperatures is not
associated with, e.g., a ferromagnetic transition, as it is the
case in manganites \cite{manganites}. A mechanism explaining such
a behavior without invoking metal-insulator transitions is
suggested in this
paper. In what follows, we focus on two of the materials with a maximum in $%
\rho _{c}$-- Sr$_{2}$RuO$_{4}$ and $\kappa $-(BEDT-TTF)$_{2}$Cu(SCN)$_{2}$%
--and show that the data for these compounds can be fitted with our model

As we have shown above, both the phonon-and electron-assisted
mechanisms mechanism lead to an increase of the tunneling
conductivity; the differences become important either at low
enough or high enough temperatures. Given the uncertainty in other
parameters of the model, we perform the fit only for the
electron-phonon mechanism. The low-temperature resistivity is
dominated by ordinary Boltzmann transport and, in principle, may
be calculated microscopically for a given Fermi surface. However,
a large number of unknown quantities, such matrix elements for
scattering and a complicated band structure, make such an approach
impractical.
\begin{figure}[th]
\vspace{1cm} \hspace{0.3cm}
\includegraphics[angle=0,width
=0.5\textwidth]
{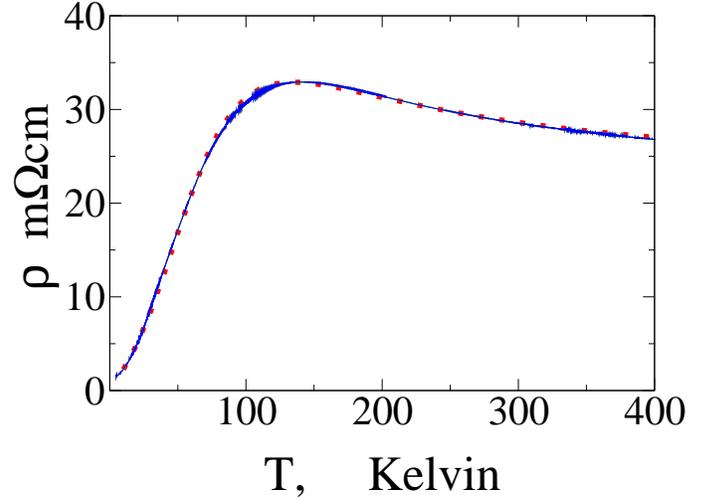} 
\hspace{2cm}
\par
\vspace{1cm}
\par
\hspace{-0.3cm} 
\includegraphics[angle=0,width
=0.5\textwidth]{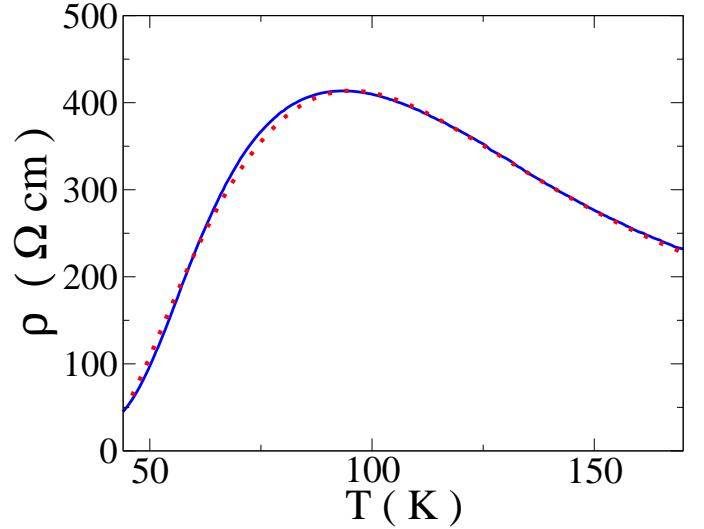} 
\caption{$\protect\rho _{c}$ vs temperature. Solid: experimental data;
dashed: fit into the phonon-assisted tunneling model. Top: Sr$_{2}$RuO$_{4}$%
. Fitting parameters: ${\protect\sigma _{\text{el}}=47\cdot
10^{3}\,\Omega
^{-1}}$ cm$^{-1}$, $\protect\omega _{D}=57$ K and $\protect\lambda =18.5$. Bottom: $\protect\kappa $%
-(BEDT-TTF)$_{2}$-Cu(SCN)$_{2}$. Fitting parameters: $\protect\sigma _{\text{%
el}}=1.5\;\Omega ^{-1}$ cm$^{-1}$, $\protect\omega _{D}=140$ K, $\protect%
\lambda =18.9$.}
\label{fig:fig2}
\end{figure}
Instead, we extract the band part of the resistivity, $\rho
_{c}^{B},$ from the low-temperature part of the experimental data.
We use the intervals of strictly metallic $T$-dependences between
10 and 50 K for Sr$_{2}$RuO$_{4}$ and between 40 and 75 K for
$\kappa $-(BEDT-TTF)$_{2}$Cu(SCN)$_{2}$), and then extrapolate the
extracted dependence of $\rho _{c}^{B}$ to higher temperatures.
The resonant part of the conductivity is calculated numerically
using Eq.(\ref{cc5}). The fit to the data for Sr$_{2}$RuO$_{4}$
and $\kappa $-(BEDT-TTF)$_{2}$Cu(SCN)$_{2}$ is shown in the top
(bottom) panels of Fig.~\ref{fig:fig2}. The quality of the fit and
reasonable values of the parameters suggest that the
phonon-assisted model is a viable mechanism of
the $c$-axis anomaly at least in these compounds. The data for Sr$_{2}$RuO$%
_{4}$ shows a tendency to saturation for $T>400$ K, which is expected for
the electron-phonon mechanism. As the electron-assisted mechanism--not
included in the fit--would lead to a further decrease in the resistivity, we
can only speculate that effective coupling for this mechanism is very small
in this material; this can be related to a rather large distance between the
planes. At temperatures above $700$ K (not shown in Fig. \ref{fig:fig2}),
the resistivity starts to rise slowly again. Although a re-entrant metallic
behavior is not explained directly by our model, it can be understood if one
recalls that for temperatures above the bandwidth of the resonant levels,
the impurity band can be effectively replaced by a single level. As we have
explained in Sec.\ref{sec:discussion},  in a single-level model the thermally activated
resonant-tunneling conductivity is insulating-like for temperatures below
the energy difference between the resonant level and metallic-like for
higher temperatures.

\section{Summary}
\label{conclusions} In this work, we suggested an explanation of
the non-metallic temperature dependence of resistivity, observed
in various layered metals. This explanation is based on the
interplay between two conduction channels of transport: the band
one with metallic-like temperature dependence and the
resonant-tunneling one with the insulating-like temperature
dependence. We developed a theory of electron-assisted tunneling
that complements the previously known theory of phonon-assisted
tunneling. According to our picture, the low-temperature part of
the resistivity is determined by relaxation of quasi-momentum due
to the electron-electron and electron-phonon interaction. At
higher temperatures, the electron-assisted and phonon-assisted
mechanisms increase the probability of resonant transmission and
lead to a decrease of resistivity with temperature.

Our model relies heavily on the assumption that resonant sites are
indeed present in real materials. Although there is an evidence
that inter-plane disorder does enhance the $c$-axis conductivity
\cite{Analytis}, a more direct verification of the resonant-level
hypothesis is needed at the moment. A combination of experimental
techniques, in which disorder is introduced controllably and
detected spectroscopically, with first-principle computational
techniques should help to resolve the issue. Such an approach
which combines low-dosage intercalation of graphite with
first-principle calculations of the energy levels of intercalated
impurities is currently being pursued
\cite{hebard_unpub},\cite{cheng}.


This research was supported by NSF-DMR-0308377. We acknowledge stimulating
discussions with B. Altshuler, A. Chubukov, A. Hebard, S. Hill, P.
Hirschfeld, P. Littlewood, D. Khmelnistkii, N. Kumar, Yu. Makhlin, A.
Mirlin, M. Reizer, A. Schofield, S. Tongay, A.A. Varlamov, and P.
W\"{o}lfle. We are indebted to A. Hebard, A. Mackenzie, and S. Tongay for
making their data available to us.

\appendix

\section{Derivation of Eq.\ref{cc7}}
\label{sec:app}
\label{phonon_assisted} At low temperatures $T\ll \omega _{D},$ the upper
limit of the integration over frequency in second term of  Eq.(\ref
{res_tun_a}) can be extended ro infinity
\begin{eqnarray}
&&f(t)=\int_{0}^{\infty }\!d\omega \frac{\omega }{\omega _{D}^{2}}\bigg[%
(1-\cos (\omega t))\left( \coth \frac{\omega }{2T}-1\right) \bigg]+  \notag
\\
&&+\int_{0}^{\omega _{D}}\frac{d\omega }{\omega _{D}^{2}}\omega \left(
1-e^{i\omega t}\right) .
\end{eqnarray}
For $T\ll \omega _{D}/\sqrt{\lambda },$ one can expand the exponential in
Eq.(\ref{sigmares_b}) as follows
\begin{equation*}
\frac{\sigma _{\mathrm{res}}}{\sigma _{\mathrm{el}}}=e^{-\lambda /2}\int dt%
\frac{i\pi tT^{2}}{\sinh ^{2}(\pi Tt+i0)}\bigg[1+\lambda \int_{0}^{\omega
_{D}}d\omega \omega e^{i\omega t}\bigg].
\end{equation*}

Using identities
\begin{eqnarray}
&& \int_{-\infty}^\infty \frac{x}{\sinh^2(\pi x +i 0)}dx=-\frac{i}{\pi}
\,\,\,\, \mathrm{and} \\
&& \int_{-\infty}^\infty\frac{xe^{ipx}dx}{\sinh^2(\pi x+i0)}=\frac{i}{\pi}%
\left( \coth\left(\frac{p}{2}\right)-1-\frac{p}{2}\sinh^{-2}\left(\frac{p}{2}%
\right)\right)  \notag
\end{eqnarray}

we get
\begin{equation*}
\hspace{-0.cm}\frac{\sigma _{\mathrm{res}}}{\sigma _{\mathrm{el}}}%
=e^{-\lambda /2}\bigg[1\!-\!\frac{\lambda }{\omega _{D}^{2}}\int_{0}^{\omega
_{D}}\!d\omega \omega \left( \coth \frac{\omega }{2T}\!-\!\frac{\omega }{2T}%
\sinh ^{-2}\frac{\omega }{2T}\!-\!1\right) \bigg].
\end{equation*}
Performing integration over frequency, we derive the low-temperature
asymptotics of $\sigma _{\mathrm{res}}$
\begin{equation*}
\hspace{-0.5cm}\frac{\sigma _{\mathrm{res}}}{\sigma _{\mathrm{el}}}%
=e^{-\lambda /2}\bigg[1+\frac{\pi ^{2}\lambda T^{2}}{3\omega _{D}^{2}}\bigg]%
\,.
\end{equation*}
Next, we consider the case of $\omega _{D}/\sqrt{\lambda }\ll T\ll \omega
_{D}$. Performing integration over frequency, we find an explicit formula
\begin{eqnarray}
&&f(t)=\frac{1}{2}+\frac{\pi ^{2}T^{2}}{3\omega _{D}^{2}}-\frac{1}{\omega
_{D}^{2}t^{2}}+\frac{\pi ^{2}T^{2}}{\omega _{D}^{2}\sinh ^{2}(\pi Tt)}+
\notag  \label{d2} \\
&&\frac{i}{\omega _{D}t}e^{i\omega _{D}t}+\frac{1-e^{i\omega _{D}t}}{\omega
_{D}^{2}t^{2}}
\end{eqnarray}
The time integration in Eq.(\ref{sigmares_b}) can be performed in the
saddle point approximation, where the saddle point solution is to be found
from
\begin{equation}
-\lambda f^{\prime }(t)-2\pi T\coth (\pi Tt)+\frac{1}{t}=0\,.  \label{eq}
\end{equation}
Recalling that $T\ll \omega _{D}$, Eq.)\ref{eq}) can be  simplified further
\begin{equation}
\label{d3}
\frac{2\cot (y)}{\sin ^{2}(y)}-\frac{\omega ^{2}}{\pi T^{2}\lambda y}+\frac{2%
}{\lambda }\left( \frac{\omega _{D}}{\pi T}\right) ^{2}\cot (y)=0,
\end{equation}
where $y=-i\pi Tt$. 
Solving Eq. (\ref{d3}) to leading order in $\omega _{D}/T,$ we find
\begin{equation*}
t^{\ast }=\frac{i}{2T}.
\end{equation*}
As a result,  we find that for $\omega _{D}/\sqrt{\lambda }\ll T\ll \omega
_{D}$ the resonant tunneling conductivity is given by
\begin{equation*}
\frac{\sigma _{\mathrm{res}}}{\sigma _{\mathrm{el}}}=\exp \left( -\frac{%
\lambda }{2}+\frac{\lambda }{\lambda }\left( \frac{\pi T}{\omega _{D}}%
\right) ^{2}\right) \,.
\end{equation*}

For high temperatures ($T\gg \omega _{D}$), the function $f(t)$ can be
approximated by
\begin{equation}
f(t)\simeq \frac{\omega _{D}}{3}(t^{2}T-it).
\end{equation}
The resonant conductivity in this temperature range is given by
\begin{equation}
\frac{\sigma _{\mathrm{res}}}{\sigma _{\mathrm{el}}}=\int dt\frac{i\pi t}{%
\sinh ^{2}(\pi t+i0)}\exp \left( -\frac{\lambda \omega _{D}}{3T}%
(t^{2}-it)\right) .  \label{aa1}
\end{equation}
For $\lambda \omega _{D}/T\gg 1$, the integral in Eq. (\ref{aa1}) is
evaluated by the saddle point approximation, yielding
\begin{equation}
\frac{\sigma _{\mathrm{res}}}{\sigma _{\mathrm{el}}}=\pi \sqrt{\frac{3\pi T}{%
4\lambda \omega _{D}}}\exp \left( -\frac{\lambda \omega _{D}}{12T}\right) \,.
\end{equation}
For $\lambda \omega _{D}/T\ll 1$, the exponent in Eq.(\ref{aa1}) can be
expanded leading to
\begin{eqnarray}
&&\frac{\sigma _{\mathrm{res}}}{\sigma _{\mathrm{el}}}=\int dt\frac{i\pi t}{%
\sinh ^{2}(\pi t+i0)}\left( 1-\frac{\lambda \omega _{D}}{3T}%
(t^{2}-it)\right) =  \notag \\
&=&1-\frac{\lambda \omega _{D}}{9T}.
\end{eqnarray}


\begin{thebibliography}{99}
\bibitem{ginsberg}  S. L. Cooper and K. E. Gray, 
in \emph{Physical Properties of High Temperature Superconductors, }edited by%
\emph{\ }D. M. Ginsberg, (World Scientific, Singapore, 1994), p. 61.

\bibitem{Terasaki}  I. Terasaki, Y. Sasago, and K. Uchinokura, Phys. Rev. B
\textbf{56}, R12685 (1997).

\bibitem{Loureiro}  S. M. Loureiro, D. P. Young, R. J. Cava, R. Jin, Y. Liu,
P. Bordet, Y. Qin, H. Zandbergen, M. Godinho, M. N\'{u}\~{n}ez-Regueiro, and
B. Batlogg, Phys. Rev. B \textbf{63}, 094109 (2001).

\bibitem{Tsukada}  I. Tsukada T. Yamamoto, M. Takagi, T. Tsubone, S. Konno,
K. Uchinokura J Phys Soc Jpn. \textbf{70}, 834 (2001), cond-matt/0012395.

\bibitem{Valla}  T. Valla, P. D. Johnson, Z. Yusof, B. Wells, Q. Li, S. M.
Loureiro, R. J. Cava, M. Mikami, Y. Mori, M. Yoshimura, T. Sasaki, Nature
\textbf{417}, 627 (2002).

\bibitem{maeno94}  Y. Maeno, H. Hashimoto, K. Yoshida, S. Nishizaki, T.
Fujita, J. G. Bednorz, and F. Lichtenberg, Nature \textbf{372}, 532 (1994).

\bibitem{hussey}  N. E. Hussey et al., Phys. Rev. B \textbf{57}, 5505 (1998).

\bibitem{srruo}  A. W. Tyler, A. P. Mackenzie, S. NishiZaki, and Y. Maeno,
Phys. Rev. B \textbf{58}, R10107 (1998).

\bibitem{frindt}  W. J. Wattamaniuk, J. P. Tidman, and R. F. Frindt,
Phys. Rev. Lett. \textbf{35,} 62 (1975).

\bibitem{graphite}  see N. B. Brandt, S. M. Chudinov, and Ya. G. Ponomarev,
\emph{Semimetals: I. Graphite and its compounds,} (North-Holland, Amsterdam,
1988) and references therein.

\bibitem{organics}  J. Singleton and C. Mielke, Contemp. Phys. \textbf{43},
63 (2002).

\bibitem{kumar}  N. Kumar and A. M. Jayannavar,
Phys. Rev. B \textbf{45}, 5001 (1992).

\bibitem{anderson_criterion}  P. W. Anderson, \emph{The theory of
superconductivity in the high T}$_{c}$ \emph{cuprates }(Princetion
University Press, 1997), p. 50.

\bibitem{dupuis}  N. Dupuis,
Phys. Rev B \textbf{56}, 9377 (1997).

\bibitem{abrikosov_loc}  Abrikosov WL

\bibitem{woelfle}  P. W\"{o}lfle and R. N. Bhatt, \emph{Electron
localization in anisotropic systems, } Phys. Rev. B\textbf{\ 30}, 3542
(1984).


\bibitem{GM}  D. B. Gutman and D. L. Maslov, Phys. Rev. Lett. (in press);
arXiv:0704.0613.

\bibitem{prange}  R. E. Prange and L. P. Kadanoff, Phys. Rev. \textbf{134},
A566 (1964).

\bibitem{singleton}  J. Singleton, P. A. Goddard, A. Ardavan, A. I. Coldea,
S. J. Blundell, R. D. McDonald, S. Tozer, and J. A. Schlueter, Phys. Rev.
Lett. \textbf{99}, 027004 (2007).

\bibitem{ioffe}  L.B. Ioffe, A.I. Larkin, A.A. Varlamov, and L. Yu,
Phys. Rev. B \textbf{47}, 8936 (1993).

\bibitem{mckenzie}  P. Moses and R. H. McKenzie,
Phys. Rev. B \textbf{60}, 7998 (1999).

\bibitem{Millis}  A. Millis, Nature \textbf{417} 599 (2002).

\bibitem{xudu}  Xu Du, S.-W. Tsai, D. L. Maslov, and A. F. Hebard, Phys.
Rev. Lett. \textbf{94}, 166601 (2005).

\bibitem{comment_1}  Experiment \cite{xudu} shows that the temperature
dependence of the transport scattering rate in graphite obeys $1/\tau _{%
\text{tr}}=0.065T$ for $25$ K$\leq T\leq 200$ K. An excellent
agreement of this result with the theory of the electron-phonon
interaction indicates that the scattering mechanism is
quasielastic scattering on phonons, when the quasiparticle
lifetime, $\tau_q$, and transport time are the same. At
$T=T_{M}=40$ K, we then have $T_{M}\tau _{q}\approx 15.$ Notice
also that since scattering is quasielastic, the Boltzmann equation
would not break down for graphite even if the product $T_{M}\tau
_{q}$ were less than one.

\bibitem{polaron_mckenzie}  U. Lundin and R. H. McKenzie,%
Phys. Rev. B \textbf{68}, 081101(R) (2003).

\bibitem{polaron_schofield}  A. F. Ho and A. J. Schofield, 
Phys. Rev. B \textbf{71}, 045101 (2005)

\bibitem{elastic}  J. Paglione, C. Lupien, W.A. MacFarlane, J.M. Perz, L.
Taillefer, Z.Q. Mao, and Y. Maeno,
\prb 65, 220506(R).

\bibitem{Turlakov}  M. Turlakov and A. J. Leggett, 
Phys. Rev. B \textbf{63}, 064518 (2001).

\bibitem{sauls}  M. J. Graf, M. Palumbo, D. Rainer, and J. A. Sauls,
Phys. Rev. B \textbf{52}, 10588 (1995).

\bibitem{levin}  A.G. Rojo and K. Levin, 
Phys. Rev. B \textbf{48}, 16861 (1993).

\bibitem{uher}  C. Uher, R. L. Hockey, and E. Ben-Jacob, Phys. Rev. B \textbf{%
35}, 4483 (1987).

\bibitem{abrikosov_res}  A. A. Abrikosov, 
Physica C \textbf{317-318, }154 (1999).

\bibitem{AA-review}  B.~L. Altshuler and A.~G. Aronov, in \emph{%
Electron--Electron Interaction In Disordered Systems}, ed. by A.L.~Efros and
M.~Pollak (Elsevier, 1985), p.1.

\bibitem{Analytis}  J. G. Analytis, A. Ardavan, S.J. Blundell, R. L. Owen,
E. F. Garman, C. Jeynes and B. J. Powell, Phys. Rev. Lett. \textbf{96},
177002 (2006).

\bibitem{Glazman_1988}  L.I. Glazman and R.I. Shekhter, Sov. Phys. JETP
\textbf{61} 163, (1988).

\bibitem{wingreen}  N. S. Wingreen, K. W. Jacobsen, and J. W. Wilkins, Phys.
Rev. Lett. \textbf{61}, 1396 (1988); Phys. Rev. B \textbf{40}, 11834 (1989).

\bibitem{Glazman_Raikh}  L.I. Glazman and M.E. Raikh, JETP Lett. \textbf{47}%
, 452 (1988).

\bibitem{Ng}  T. K. Ng and P.A. Lee, Phys. Rev. Lett. \textbf{61}, 1768
(1988).

\bibitem{Larkin_Matveev}  A.I. Larkin and K. Matveeev, Sov. Phys. JETP
\textbf{66}, 580 (1987).

\bibitem{Geim}  A. K. Geim, P. C. Main, N. La Scala, L. Eaves, T. J. Foster,
P. H. Beton, J. W. Sakai, F. W. Sheard, M. Henini, G. Hill, and M. A. Pate,
Phys. Rev. Lett. \textbf{72} 2061 (1994).

\bibitem{Glazman_Matveev}  L.I. Glazman and K.A. Matveev, Sov. Phys. JETP
\textbf{67} 1276 (1988).

\bibitem{Beasley}  D. Ephron, M.R. Beasley, H. Bahlouli, K.A. Matveev
Phy.Rev. B \textbf{49} 2989 (1994); H. Bahlouli, K.A. Matveev, D. Ephron,
M.R. Beasley Phy.Rev. B \textbf{49} 14496 (1994).

\bibitem{CL}  A.O. Caldeira, A. J. Leggett, Phys. Rev. Lett. \textbf{46},
211 (1981).

\bibitem{Lipkin}  H.J. Lipkin, {\it Quantum mechanics; new approaches to
selected topics}, (North-Holland, New York, 1973).

\bibitem{lang_firsov}  I. G. Lang and Yu. A. Firsov, Sov. Phys. JETP \textbf{%
16}, 1301 (1961); Sov. Phys. Solid State \textbf{5}, 2049 (1964).

\bibitem{Mahan}  G. D. Mahan {\it Many-particle physics} (Plenum Press, New
York, 1981).

\bibitem{Finkelstein83}  A.~M.~Finkel'stein, Sov. Phys. JETP \textbf{57}, 97
(1983); \textit{ibid} \textbf{59}, 212 (1984); Sov. Sci. Rev. A \textbf{14},
1 (1990).

\bibitem{hebard_unpub}  S. Tongay and A. F. Hebard, private communication.

























\bibitem{ioffe_millis}  L. B. Ioffe and A. J. Millis, Phys. Rev. B \textbf{58%
}, 11631 (1998); \emph{ibid.} \textbf{61}, 9077 (2000).

\bibitem{rudin}  A. M. Rudin, I. L. Aleiner, and L. I. Glazman, Phys. Rev. B
\textbf{55}, 9322 (1997).

\bibitem{zna}  G. Zala, B. N. Narozhny, and I. L. Aleiner, Phys. Rev. B
\textbf{64}, 214204 (2001)

\bibitem{reizer}  D. V. Khveshchenko and M. Reizer, Phys. Rev. B \textbf{57}%
, R4245 (1998).

\bibitem{Mishchenko_2001}  E. G. Mishchenko, A. V. Andreev Phys. Rev. B
\textbf{65}, 235310 (2002).

\bibitem{manganites} M. B. Salamon and M. Jaime,  Rev. Mod. Phys. {\bf 73}, 583 (2001).

\bibitem{cheng}  H.-P. Cheng, private communication.
\end{thebibliography}
\end{document}